\documentclass[12pt]{article}   
\usepackage{epsfig}
\newcommand{\mysection}{\setcounter{equation}{0}\section}

\def\beq{\begin{equation}}
\def\eeq{\end{equation}}
\def\beqa{\begin{eqnarray}}
\def\eeqa{\end{eqnarray}}

\newlength{\dinwidth} \newlength{\dinmargin}
\begin{document}

\begin{center}
{\Large \bf $W$ production at large transverse momentum 
at the Large Hadron Collider}
\end{center}
\vspace{2mm}
\begin{center}
{\large Richard J. Gonsalves,$^a$ Nikolaos Kidonakis,$^b$ 
and Agust{\' \i}n Sabio Vera$^c$}\\
\vspace{2mm}
{\it $^a$ Department of Physics, University at Buffalo,
The State University of New York\\
Buffalo, NY 14260-1500\\  
\vspace{2mm}
$^b$ Kennesaw State University, Physics \#1202 \\
1000 Chastain Rd., Kennesaw, GA 30144-5591, USA\\
\vspace{2mm}
$^c$ II. Institut f{\"u}r Theoretische Physik, Universit{\"a}t Hamburg\\
Luruper Chaussee 149, 22761~Hamburg, Germany}
\end{center}

\begin{abstract}
We study the production of $W$ bosons at large transverse momentum
in $pp$ collisions at the Large Hadron Collider (LHC). 
We calculate the complete next-to-leading 
order (NLO) corrections to the differential cross section.
We find that the NLO corrections provide a large increase to the cross section
but, surprisingly, do not reduce the scale dependence
relative to leading order (LO). We also calculate
next-to-next-to-leading-order (NNLO) soft-gluon corrections and find that,  
although they are small, they significantly reduce the scale dependence 
thus providing a more stable theoretical prediction.
\end{abstract}

\thispagestyle{empty} \newpage \setcounter{page}{2}

\mysection{Introduction}

The study of $W$-boson production in hadron colliders is important in
testing the Standard Model \cite{pdg} and in estimating backgrounds to
Higgs production and new physics \cite{back}.
Precise theoretical predictions for $W$ production at the Large Hadron
Collider (LHC) are needed to exploit fully the large number of such
events that are anticipated when the LHC begins operation in 2007.
Accurate predictions for this process, which has a particularly clean
experimental signature when the $W$ decays to leptons, are needed to
reduce the systematic uncertainties in precision measurements, e.g.,
of the $W$ mass and decay width.
Due to the large number of events expected, this process can also be
used to determine the parton-parton luminosity \cite{lum}.
Furthermore, accurate calculations for $W$ production at large
transverse momentum, $Q_T$, are required to distinguish the Standard
Model prediction from signals of possible new physics, such as new
gauge bosons and extra dimensions \cite{new}, which may be expected to
enhance the $Q_T$ distribution at large $Q_T$.

Calculations of the next-to-leading order (NLO) 
cross section for $W$ production at large 
transverse momentum at the Fermilab Tevatron collider were 
presented in Refs. \cite{AR,gpw}.
These predictions have been tested experimentally by the CDF and
D\O\ collaborations \cite{cdfd0}, and found to be consistent with the
data at large $Q_T$.
The NLO corrections contribute to enhance the
differential distributions in $Q_T$ of the $W$ boson 
and they reduce the factorization and renormalization
scale dependence of the cross section.
A recent study \cite{NKASV} included soft-gluon corrections
through next-to-next-to-leading order (NNLO),
which provide additional enhancements and a further reduction
of the scale dependence.
 
At leading order (LO) in the Quantum Chromodynamic (QCD) coupling
$\alpha_s$, a $W$-boson can be produced at large $Q_T$ by recoiling
against a single parton which decays into a jet of hadrons.
The NLO corrections to this cross section involve one-loop parton
processes with a virtual gluon, and real radiative processes with
two partons in the final state. The virtual corrections involve
ultraviolet divergences which renormalize $\alpha_s$ and cause it to
depend on a renormalization energy scale which we will take to 
be ${\sim}Q_T$.
Both real and virtual corrections have soft and collinear divergences
which arise from the masslessness of the gluons and our approximation
of zero masses for the quarks.
The soft divergences cancel between real and virtual processes.
The net collinear singularities are factorized in a process-independent
manner and absorbed into factorization-scale-dependent parton 
distribution functions.
Complete analytic expressions for the inclusive NLO cross sections in
terms of the partonic Mandelstam variables were presented in
Refs.~\cite{AR,gpw}, and form the basis for the NLO results in this work.

The calculation of hard-scattering cross sections near partonic threshold,
such as $W$-boson
production at large transverse momentum, involves corrections from the
emission of soft gluons from the partons in the process.
At each order in perturbation theory one encounters large
logarithms that arise from the cancellations  
between real emission and virtual processes, 
due to the limited phase space available for real gluon emission
near partonic threshold. These threshold corrections 
formally exponentiate as a result of the factorization
properties \cite{NKres} of the cross section.
Expansions of the resummed cross section can be derived 
in principle to any higher order.
A unified approach to the calculation of NNLO soft-gluon 
corrections for hard-scattering 
processes was recently presented in Ref. \cite{NKuni}.
Analytic expressions for $W$ production are given in Ref. \cite{NKASV}.

We note that there has also been work on resummation of Sudakov logarithms
at small transverse momentum for electroweak boson production in hadron
colliders \cite{ERV,BY,KuSt,NY} as well as on joint resummation of $Q_T$ 
and threshold logarithms \cite{KSV}. 
The results presented in this paper for large $Q_T\ge20$ GeV are not
sensitive to this resummation.
 
In this paper we calculate the complete NLO
corrections as well as the NNLO soft-gluon corrections
for $W$-boson production at large transverse momentum at the LHC.
We study the size of the corrections and their significance in 
stabilizing the cross section versus changes in the factorization and 
renormalization scales.

\mysection{Kinematics and Partonic Cross Sections}
 
For the production of a $W$ boson, with momentum $Q$, in 
collisions of two hadrons $h_A$ and $h_B$ with momenta $P_A$ and $P_B$
\beq
h_A(P_A)+h_B(P_B) \longrightarrow W(Q) + X \, ,
\eeq
where $X$ denotes all additional particles in the final state,
we can write the factorized single-particle-inclusive cross section as
\beqa
E_Q\,\frac{d\sigma_{h_Ah_B{\rightarrow}W(Q)+X}}{d^3Q} &=&
\sum_{f_a,f_b} \int dx_a \, dx_b \; \phi_{f_a/h_A}(x_a,\mu_F^2)
\; \phi_{f_b/h_B}(x_b,\mu_F^2)
\nonumber \\ && \hspace{-10mm} \times \,
E_Q\,\frac{d\hat{\sigma}_{f_af_b{\rightarrow}W(Q)+X}}{d^3Q}
(s,t,u,Q^2,\mu_F,\mu_R,\alpha_s(\mu_R^2)) \label{factor}
\label{factW}
\eeqa
where $E_Q=Q^0$, the parton distribution $\phi_{f/h}$ is the probability
density in the momentum fraction $x$ for finding parton $f$ 
with momentum $p=xP$ in hadron $h$,
and $\hat{\sigma}$ is the perturbative parton-level cross section.
The initial-state collinear singularities are factorized into the
parton distributions at factorization scale $\mu_F$, while $\mu_R$ is the
renormalization scale.
 
At the parton level, the lowest-order subprocesses for the
production of a $W$ boson and a parton involve quarks $q$,
anti-quarks $\bar q$ and gluons $g$:
\beqa
q(p_a)+g(p_b) &\longrightarrow& W(Q) + q(p_c)  \, ,
\nonumber \\
q(p_a)+{\bar q}(p_b) &\longrightarrow& W(Q) + g(p_c)  \, .
\label{partsub}
\eeqa
The partonic kinematical invariants in the process are
$s=(p_a+p_b)^2$, $t=(p_a-Q)^2$, $u=(p_b-Q)^2$, and
$s_2 = s + t + u - Q^2=(p_a+p_b-Q)^2$. Here $s_2$ 
is the invariant mass of the system recoiling
against the electroweak boson at the parton level 
and parameterizes the inelasticity of the parton scattering,
taking the value $s_2=0$ for one-parton production.

Schematically, the parton level cross sections at NLO have the form
\beq
E_Q\,\frac{d\hat{\sigma}_{f_af_b{\rightarrow}W(Q)+X}}{d^3Q}=
\delta(s_2)\alpha_s(\mu_R^2)\left[A(s,t,u)+\alpha_s(\mu_R^2)
B(s,t,u,\mu_R)\right] + \alpha_s^2(\mu_R^2)C(s,t,u,s_2,\mu_F)\, .
\label{nloxc}
\eeq
The coefficient functions $A$, $B$ and $C$ depend on the parton flavors
$f_af_b$ and on electroweak parameters (which are suppressed),
in addition to the kinematic variables $s,t,u,s_2$, and scales
$\mu_R,\mu_F$ shown explicitly.
The coefficient $A(s,t,u)$ arises from the LO processes in Eq. (\ref{partsub}).
The functions $B$ and $C$ result from a perturbative QCD calculation of the
virtual and real corrections, respectively, to the processes (\ref{partsub}).
The NLO calculation is done using dimensional regularization and modified
minimal subtraction ($\overline{\rm MS}$) to deal with ultraviolet, soft, and
collinear divergences.
The function $B(s,t,u,\mu_R)$ is the finite (after renormalization) sum
of virtual corrections and of singular terms ${\sim}\delta(s_2)$ in the
real radiative corrections.
The function $C(s,t,u,s_2,\mu_F)$ is the finite (after factorization)
contribution from real emission processes away from the edge $s_2=0$
of phase space.
Analytic expressions for these functions from Ref.~\cite{gpw} were used
to compute the numerical NLO results in this paper.

In general, the partonic cross section $\hat{\sigma}$
includes distributions with respect
to $s_2$ at $n$-th order in the QCD coupling $\alpha_s$ of the type
\beq
\left[\frac{\ln^{m}(s_2/Q_T^2)}{s_2} \right]_+, \hspace{10mm} m\le 2n-1\, ,
\label{zplus}
\eeq
defined by their integral with any smooth function $f$ by
\beqa
\int_0^{s_{2 \, max}} ds_2 \, f(s_2) \left[\frac{\ln^m(s_2/Q_T^2)}
{s_2}\right]_{+} &\equiv&
\int_0^{s_{2\, max}} ds_2 \frac{\ln^m(s_2/Q_T^2)}{s_2} [f(s_2) - f(0)]
\nonumber \\ &&
{}+\frac{1}{m+1} \ln^{m+1}\left(\frac{s_{2\, max}}{Q_T^2}\right) f(0) \, .
\label{splus}
\eeqa
These ``plus'' distributions are the remnants of cancellations between
real and virtual contributions to the cross section.
Below we will make use of the terminology that at $n$-th order in $\alpha_s$
the leading logarithms (LL)
are those with $m=2n-1$ in Eq. (\ref{zplus}), 
the next-to-leading logarithms (NLL) are those with $m=2n-2$, 
the next-to-next-to-leading logarithms (NNLL) are those with $m=2n-3$,
and the next-to-next-to-next-to-leading logarithms (NNNLL)  
are those with $m=2n-4$.
Below the term ``NNLO-NNNLL'' means that the soft-gluon contributions 
through NNNLL to the NNLO corrections have been included and added to the
complete NLO result (for details see Ref. \cite {NKASV}).

\mysection{Numerical results}

\begin{figure}[htb] 
\setlength{\epsfxsize=0.85\textwidth}
\setlength{\epsfysize=0.55\textheight}
\centerline{\epsffile{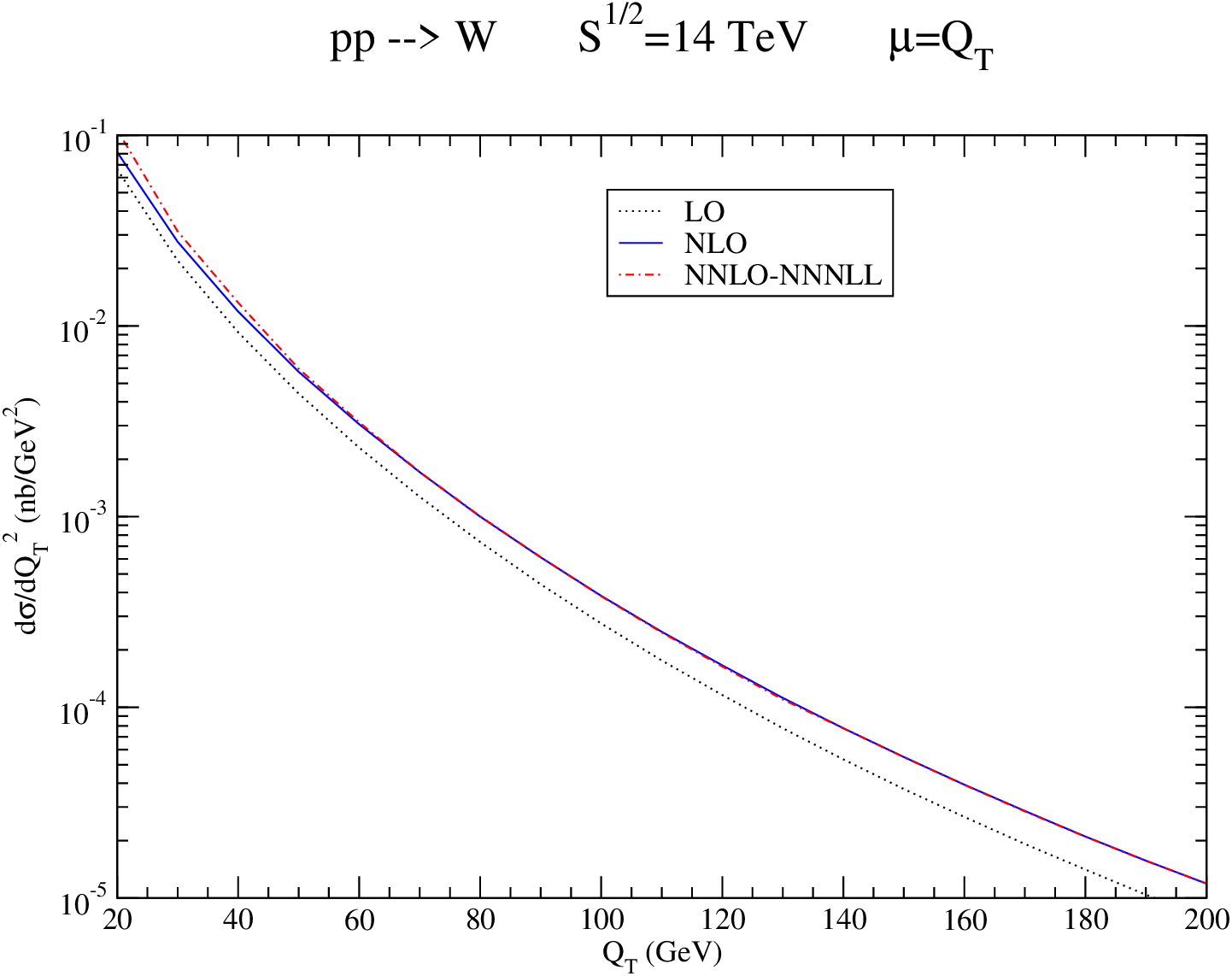}}
\caption[]{The differential cross section,
$d\sigma/dQ_T^2$, for $W$ production in $pp$ collisions
at the LHC with $\sqrt{S}=14$ TeV and $\mu=\mu_F=\mu_R=Q_T$.
Shown are the LO, NLO, and NNLO-NNNLL results.}
\label{wlhc} 
\end{figure}

\begin{figure}[htb] 
\setlength{\epsfxsize=0.85\textwidth}
\setlength{\epsfysize=0.55\textheight}
\centerline{\epsffile{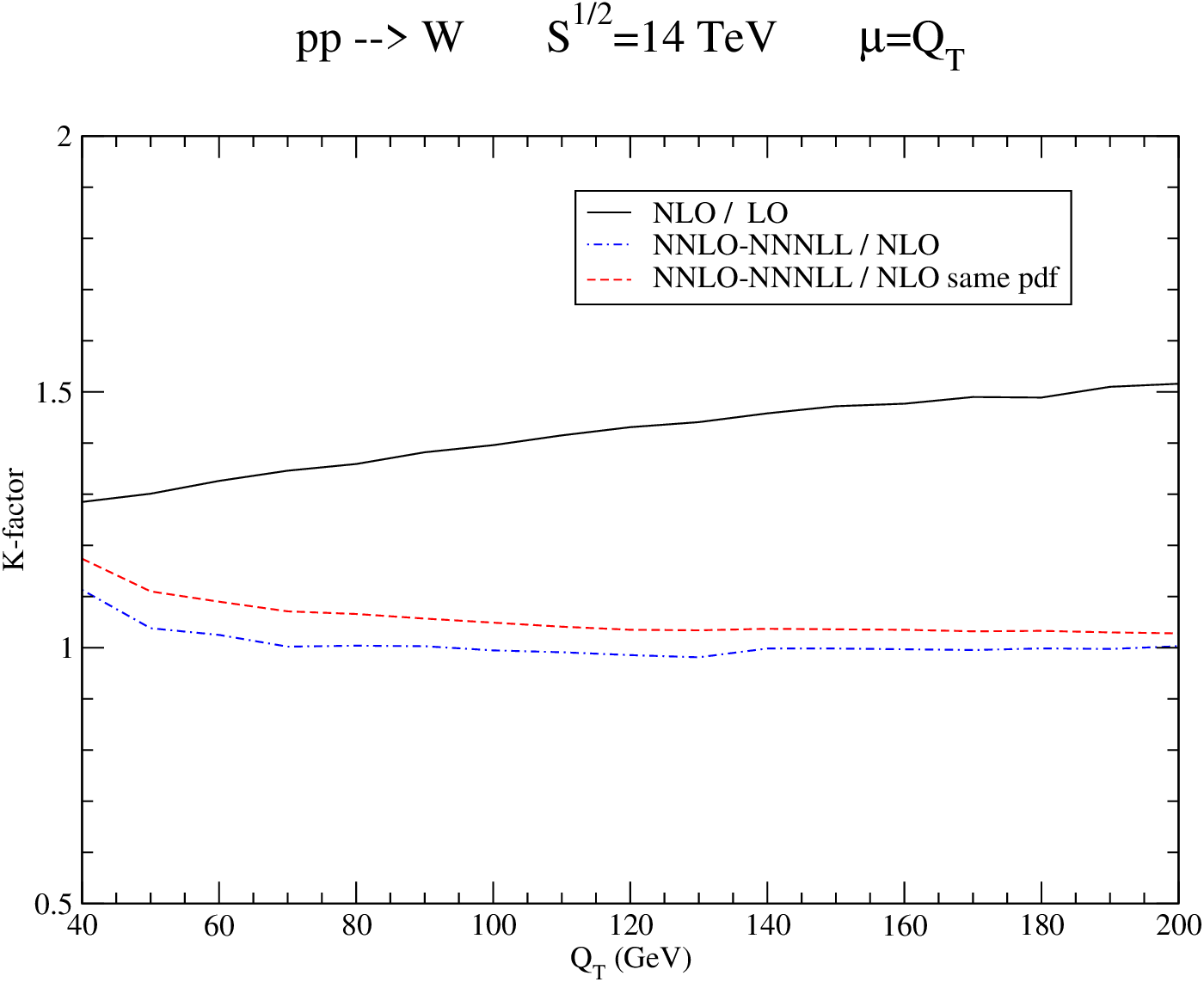}}
\caption[]{The $K$-factors for the differential cross section,
$d\sigma/dQ_T^2$,  for $W$ production 
in $pp$ collisions at the LHC with $\sqrt{S}=14$ TeV 
and $\mu=\mu_F=\mu_R=Q_T$.
Shown are the $K$- factors for NLO/LO and NNLO-NNNLL/NLO.
The latter is shown with the same or different parton distribution 
functions (pdf) for the NLO and NNLO results.
}
\label{Kwlhc} 
\end{figure}

We now apply our results to $W$ production at large transverse momentum
at the LHC. Throughout we use the MRST2002 
parton densities \cite{MRST}.
The QCD coupling $\alpha_s(\mu_R^2/\Lambda^2)$ is computed using the 
appropriate value of the 4-flavor QCD scale parameter
$\Lambda_4$ required by the LO, NLO, and NNLO parton densities,
and assuming five flavors of quarks (u,d,s,c,b).
The coupling is evolved using the massless LO, NLO and NNLO $\beta$-function
as appropriate, and using the standard \cite{pdg} effective-flavor scheme to
cross heavy-quark thresholds.
The electroweak coupling $\alpha(M_Z^2)$ is evaluated at the mass of the
$Z$ boson, and standard values \cite{pdg} are used for the various electroweak
parameters.

In all of the numerical results we present the sum of cross sections for
$W^-$ and $W^+$ production.
$W$ bosons at the LHC will be detected primarily through their leptonic
decay products e.g., $W^-\rightarrow \ell\bar\nu_\ell$.
Our cross sections should therefore be multiplied by the appropriate
branching ratios (${\sim}0.11$ for each lepton species).
Our cross sections are integrated over the full phase space of the
$W$ and final state partons.
Detailed comparisons with experiment will require realistic cuts
on lepton and jet momenta.
These modifications will affect the overall normalization of the cross
sections, but they should not affect the main results of this paper
concerning the convergence and reliability of perturbative corrections.

In Fig. \ref{wlhc} we plot the transverse momentum distribution,
$d\sigma/dQ_T^2$, for $W$ production at the LHC
with $\sqrt{S}=14$ TeV. Here we set $\mu_F=\mu_R=Q_T$ and denote this common
scale by $\mu$.
We plot LO, NLO, and NNLO-NNNLL results. 
In the LO result we use LO parton densities, in the NLO result we use
NLO parton densities, and in the NNLO-NNNLL result we use 
NNLO parton densities. We see that the
NLO corrections provide a significant enhancement of the LO $Q_T$ 
distribution. The NNLO-NNNLL corrections provide a further
rather small enhancement of the $Q_T$ distribution, which is hardly visible
in the plot.
Part of the reason for the small difference between the NLO and
the NNLO-NNNLL curves is that the NNLO parton densities
reduce the cross section.
As we will see below, the NNLO-NNNLL corrections can be much bigger for
other choices of factorization and renormalization scales.
We also note that the NNLO-NNNLL result is practically the same as
at NNLO-NNLL (i.e. if we had kept one level less of logarithms),
and that while the LL, NLL, and NNLL terms are complete, the NNNLL terms 
are not complete but include the dominant contributions (see \cite{NKASV}
for details). 

\begin{figure}[htb] 
\setlength{\epsfxsize=0.85\textwidth}
\setlength{\epsfysize=0.55\textheight}
\centerline{\epsffile{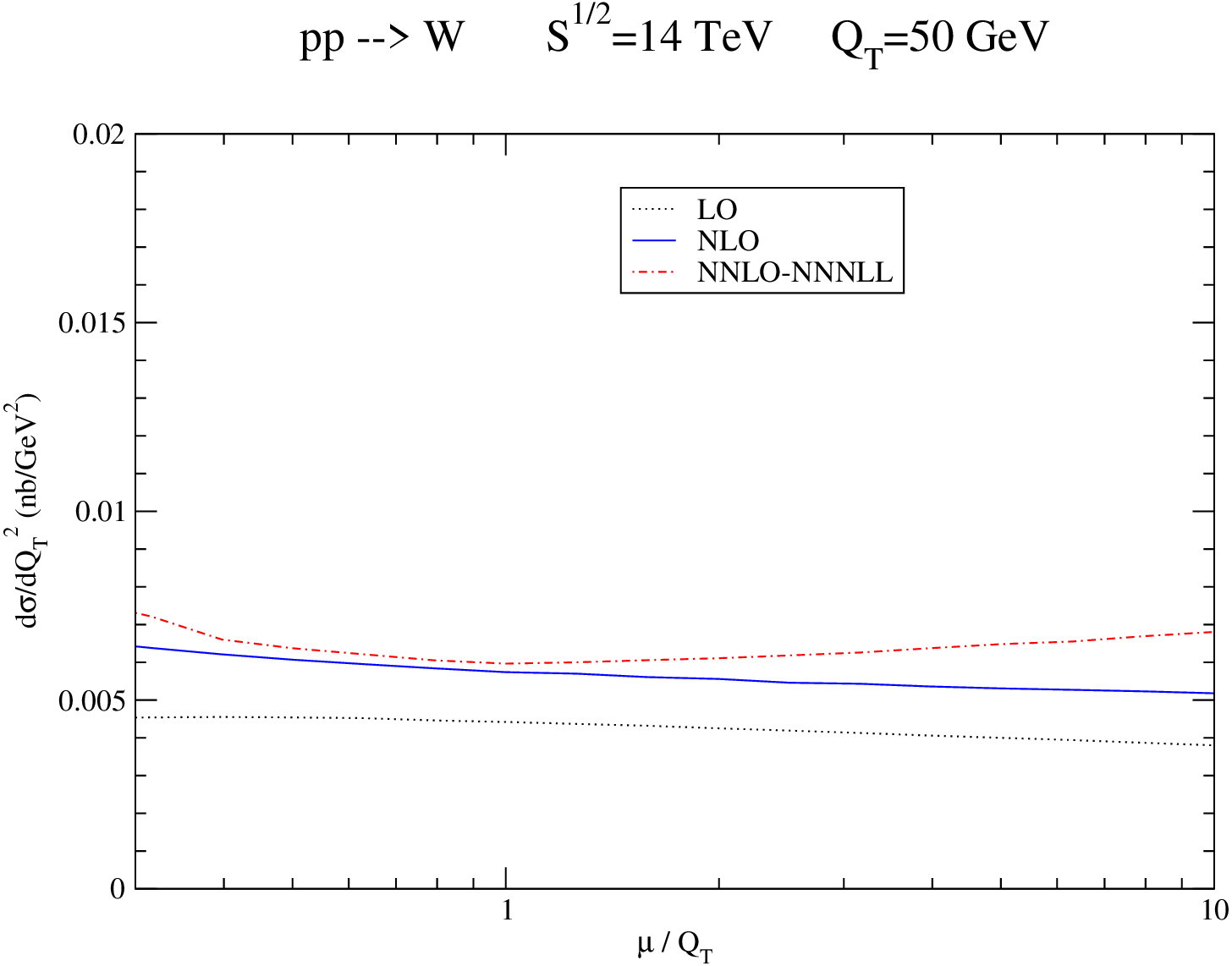}}
\caption[]{The differential cross section,
$d\sigma/dQ_T^2$, for $W$ production in $pp$ collisions
at the LHC with $\sqrt{S}=14$ TeV, $Q_T=50$ GeV, and 
$\mu=\mu_F=\mu_R$.
Shown are the LO, NLO, and NNLO-NNNLL results.
}
\label{mu50} 
\end{figure}

In Fig. \ref{Kwlhc} we plot the $K$-factors, i.e. the ratios of
cross sections at various orders and accuracies
to the LO cross section, all with $\mu_F=\mu_R=Q_T$,
in the high-$Q_T$ region.
We see that the NLO/LO ratio is rather large; 
the NLO corrections
increase the LO result by about 30\% to 50\% in the $Q_T$ range shown.
In contrast, the NNLO-NNNLL/NLO ratio for this choice of scale is rather small.
Part of the reason for this is that the NNLO parton densities are 
significantly smaller than the NLO parton densities. To make this point
more clearly we also show a curve for the NNLO-NNNLL/NLO $K$-factor where
both the NNLO-NNNLL and the NLO cross sections are calculated using the
same NNLO parton densities. The ratio is then significantly bigger.
We can also see that the NNLO-NNNLL/NLO $K$-factors are
nearly constant over the $Q_T$ range shown even though
the differential cross sections themselves span three orders of magnitude in 
this range.

\begin{figure}[htb] 
\setlength{\epsfxsize=0.85\textwidth}
\setlength{\epsfysize=0.55\textheight}
\centerline{\epsffile{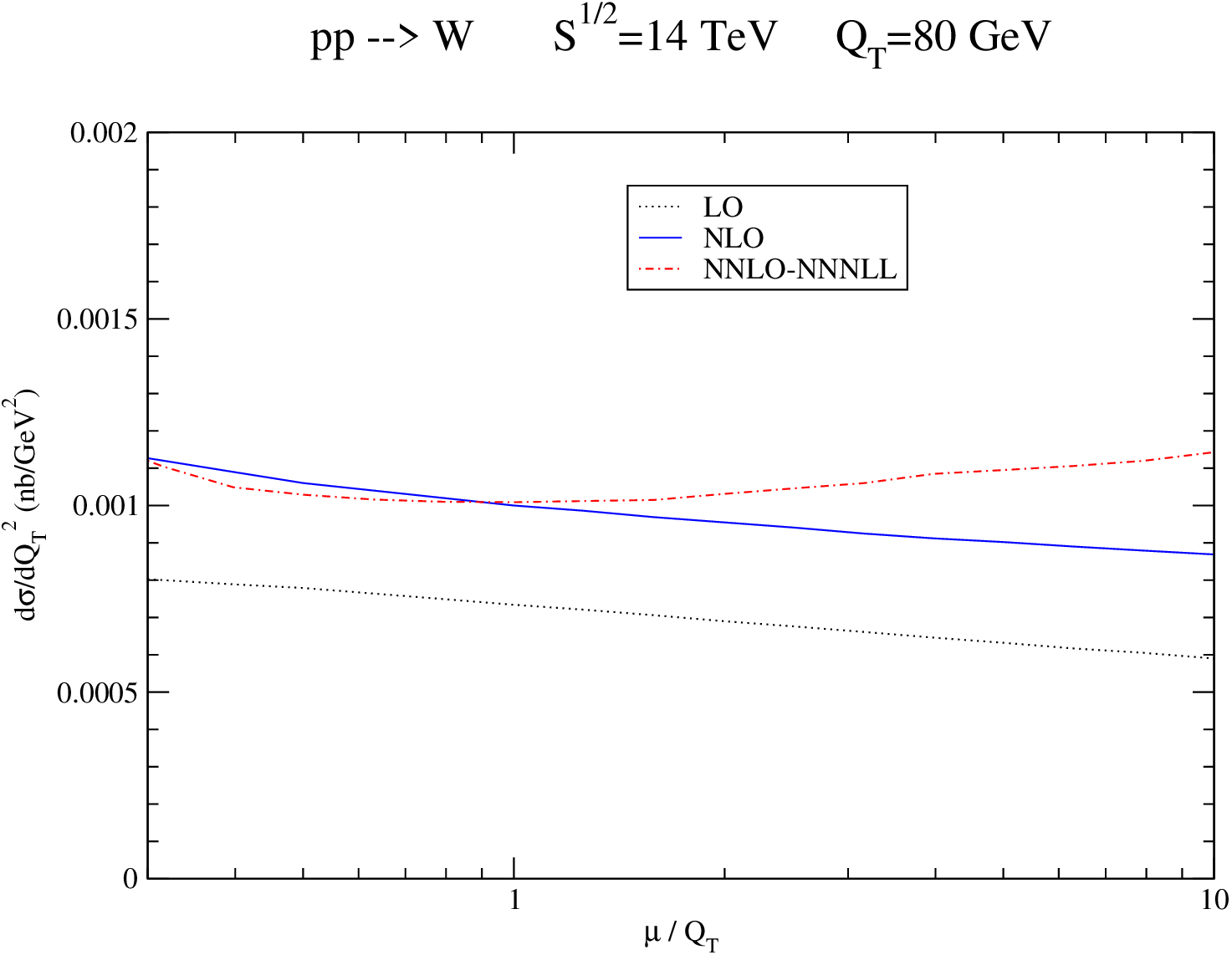}}
\caption[]{The differential cross section,
$d\sigma/dQ_T^2$, for $W$ production in $pp$ collisions
at the LHC with $\sqrt{S}=14$ TeV, $Q_T=80$ GeV, and 
$\mu=\mu_F=\mu_R$.
Shown are the LO, NLO, and NNLO-NNNLL results.
}
\label{mu80} 
\end{figure}

\begin{figure}[htb] 
\setlength{\epsfxsize=0.85\textwidth}
\setlength{\epsfysize=0.55\textheight}
\centerline{\epsffile{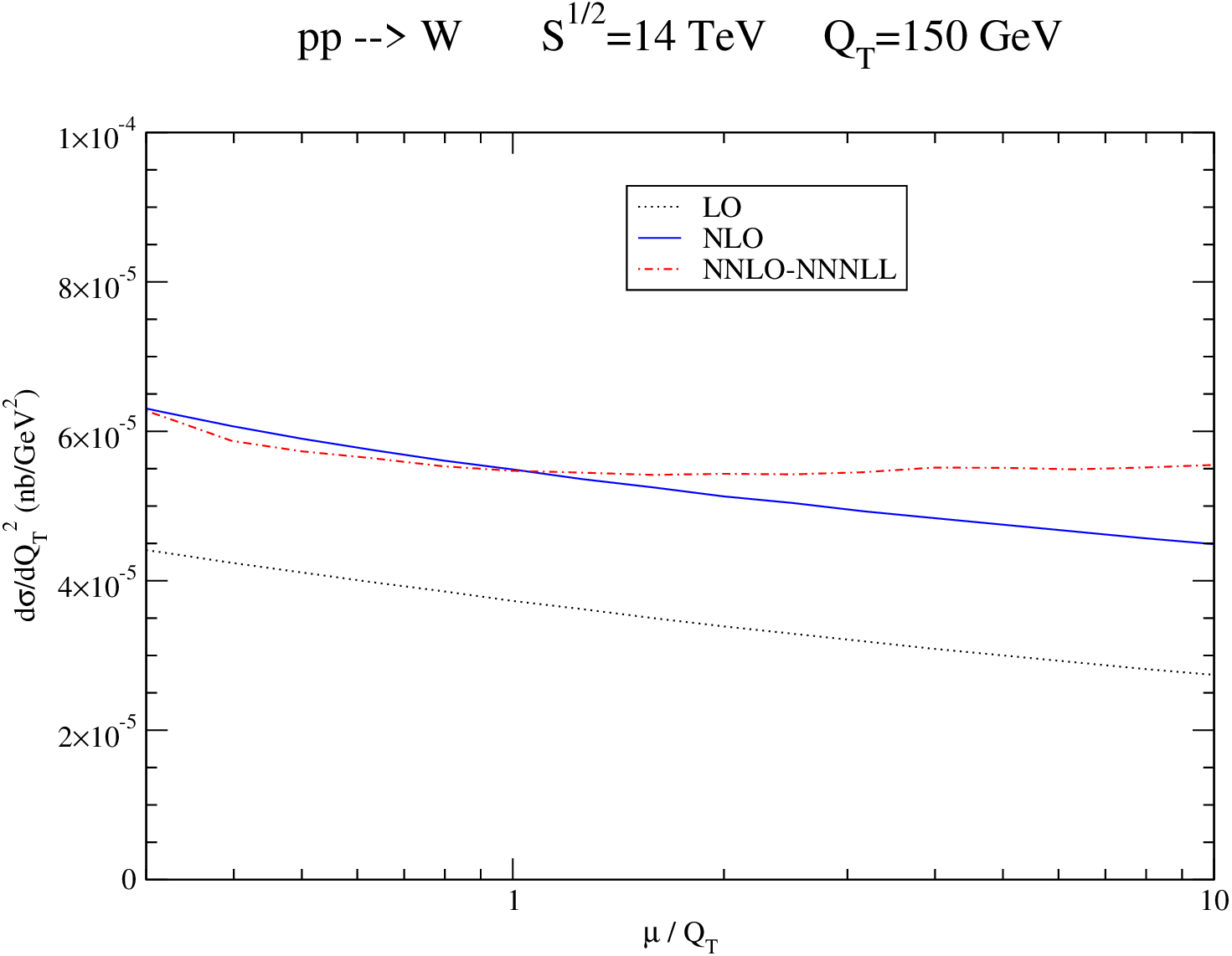}}
\caption[]{The differential cross section,
$d\sigma/dQ_T^2$, for $W$ production in $pp$ collisions
at the LHC with $\sqrt{S}=14$ TeV, $Q_T=150$ GeV, and 
$\mu=\mu_F=\mu_R$.
Shown are the LO, NLO, and NNLO-NNNLL results.
}
\label{mu150} 
\end{figure}

In Figs. \ref{mu50}, \ref{mu80}, and \ref{mu150} 
we plot the scale dependence of the differential cross section
for $Q_T=50$, 80, and 150 GeV, respectively. We set $\mu_F=\mu_R$ 
and denote this common scale by $\mu$. We plot
the differential cross section versus $\mu/Q_T$ over two
orders of magnitude. We note that the scale dependence
of the cross section is not reduced when the NLO corrections are
included, but we have an improvement when the NNLO-NNNLL corrections
are added.
The NNLO-NNNLL result approaches the scale independence expected of a truly
physical cross section.

\begin{figure}[htb] 
\setlength{\epsfxsize=0.85\textwidth}
\setlength{\epsfysize=0.55\textheight}
\centerline{\epsffile{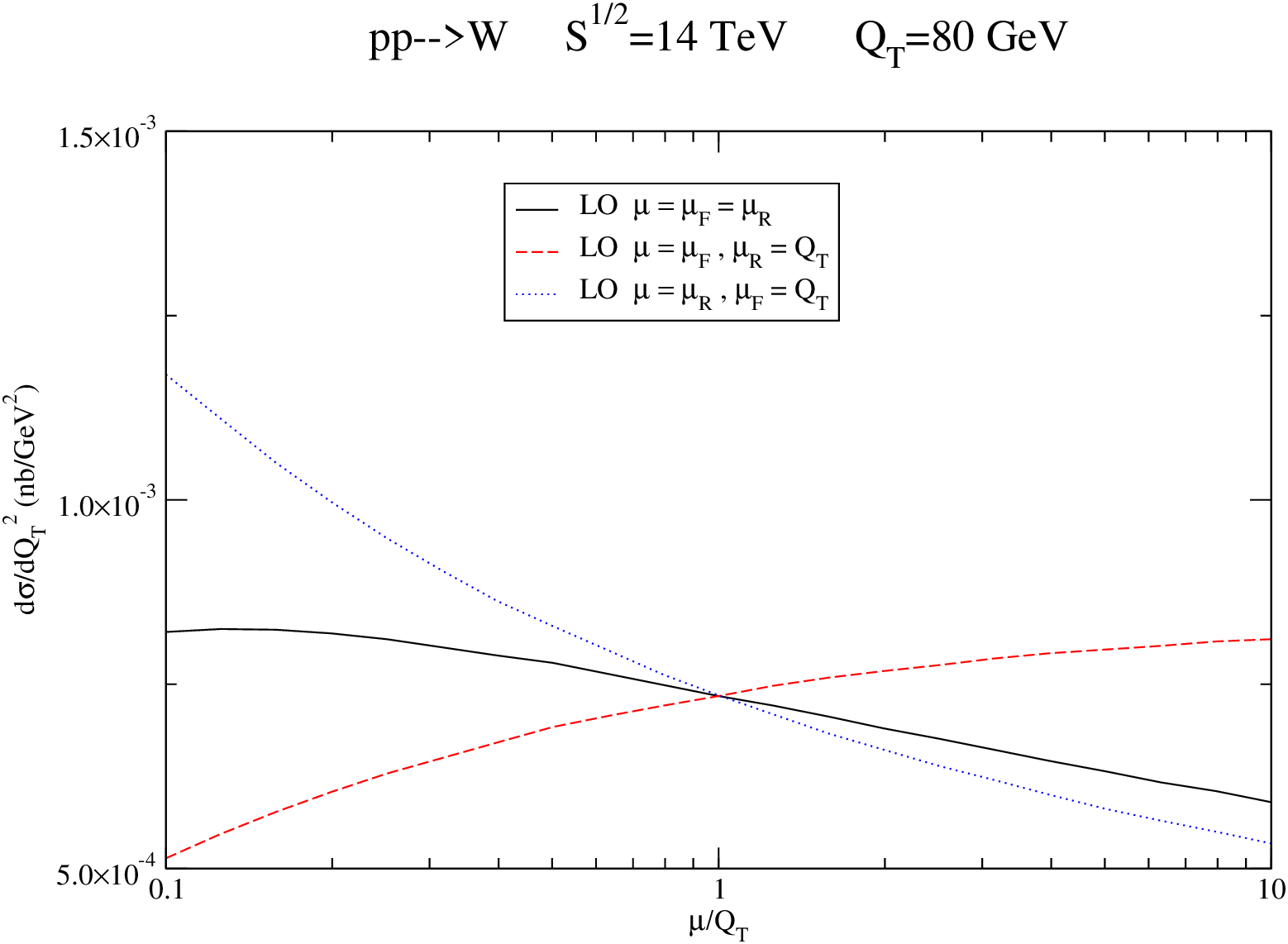}}
\caption[]{The differential cross section,
$d\sigma/dQ_T^2$, for $W$ production in $pp$ collisions
at the LHC with $\sqrt{S}=14$ TeV, $Q_T=80$ GeV.
Shown are the LO results with $\mu_F=\mu_R$, and with $\mu_F$ and
$\mu_R$ varied separately.
}
\label{mufr80} 
\end{figure}

It is interesting, and perhaps a little surprising, that the scale
dependence of the LO results in Figs. \ref{mu50}, \ref{mu80}, \ref{mu150} 
is less pronounced than that of the NLO results.
One might expect that the scale dependence should decrease with each
successive order of perturbation theory.
This is indeed what is observed at Tevatron energies \cite{gpw,NKASV},
where the LO cross section increases monotonically with decreasing scale
and has positive curvature. 
The LO curves in Figs.  \ref{mu50}, \ref{mu80}, \ref{mu150} have
little curvature, and at $Q_T=80$ GeV the curve actually turns over
at small $\mu/Q_T$.
In Fig. \ref{mufr80} we plot the LO scale dependence separately for
$\mu_F$ and $\mu_R$ with the other held fixed. 
The cross section increases with positive curvature
as the renormalization scale $\mu_R$  is decreased: 
this is the expected behavior at LO due to asymptotic freedom.
The $\mu_F$ dependence however has negative curvature and the cross section
increases with scale.
This behavior is due to the fact that the cross section is dominated by
the gluon-initiated process $qg\rightarrow Wq$.
The gluon density in the proton increases rapidly with scale at fixed $x$
smaller than ${\sim}0.01$.
At LHC energies, the $\mu_R$ and $\mu_F$ dependencies cancel one another
approximately, while the scale dependence of the LO cross section at
Tevatron energies is dominated by the $\mu_R$ dependence of $\alpha_s$.

\begin{figure}[htb] 
\setlength{\epsfxsize=0.85\textwidth}
\setlength{\epsfysize=0.55\textheight}
\centerline{\epsffile{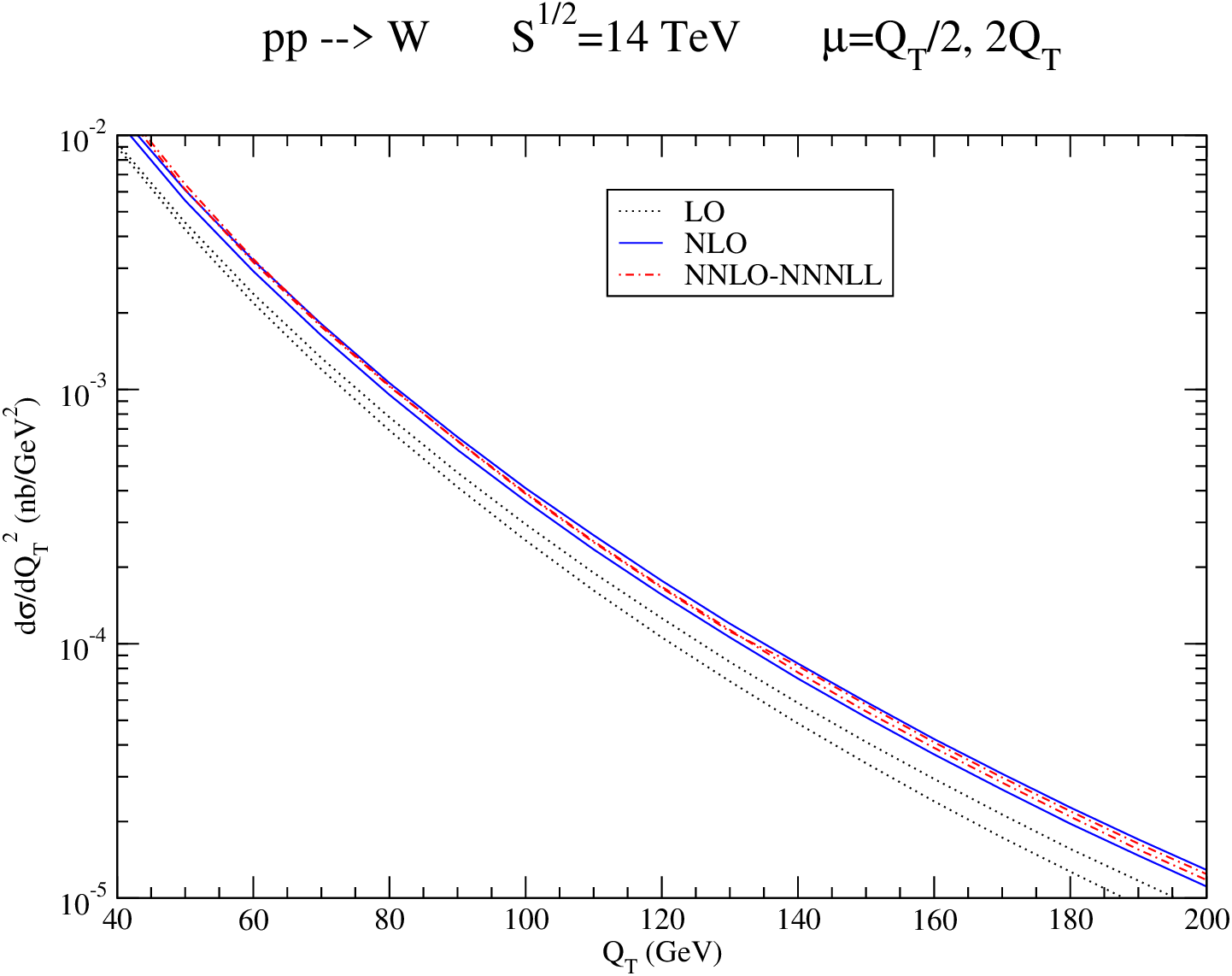}}
\caption[]{The differential cross section,
$d\sigma/dQ_T^2$, for $W$ production in $pp$ collisions
at the LHC with $\sqrt{S}=14$ TeV and 
$\mu=\mu_F=\mu_R=Q_T/2$ or $2Q_T$.
Shown are the LO, NLO, 
and NNLO-NNNLL results. The upper lines
are with $\mu=Q_T/2$, the lower lines with $\mu=2 Q_T$.
}
\label{wmu} 
\end{figure}

In Fig. \ref{wmu} we plot the differential cross section
$d\sigma/dQ_T^2$ at high $Q_T$ with $\sqrt{S}=14$ TeV
for two values of scale, $Q_T/2$ and $2Q_T$,
often used to display the uncertainty due to scale variation.
We note that while the variation of the LO cross section is
significant and the variation at NLO is similar to LO, at
NNLO-NNNLL it is very small. In fact the two NNLO-NNNLL curves
lie very close to or on top of each other. These results are consistent with
Figs. \ref{mu50}, \ref{mu80}, \ref{mu150}.

\mysection{Conclusions}
 
We have presented the full NLO and the NNLO soft-gluon corrections
for $W$ production at large transverse momentum
in $pp$ collisions at the LHC.
We have shown that the NLO corrections are large
but do not diminish the scale dependence of the cross section 
relative to LO.
The NNLO-NNNLL corrections are very small for $\mu=Q_T$
but they significantly decrease the factorization and renormalization
scale dependence of the transverse momentum distributions.
These precise and reliable perturbative QCD predictions for inclusive
$W$ production will facilitate precision tests of the Standard Model,
measurements of parton densities, and searches for new physics at
the LHC.

\end{document}